\documentclass[fleqn,usenatbib]{mnras}

\usepackage{newtxtext,newtxmath}

\usepackage[T1]{fontenc}

\DeclareRobustCommand{\VAN}[3]{#2}
\let\VANthebibliography\thebibliography
\def\thebibliography{\DeclareRobustCommand{\VAN}[3]{##3}\VANthebibliography}

\usepackage{graphicx}	
\usepackage{amsmath}	
\usepackage{amssymb}	
\usepackage{color}
\usepackage{xcolor}

\newcommand{\oiii}{[O~{\footnotesize III}]}

\title[A30]{Abell 30 -- A Binary Central Star Among the Born-Again Planetary Nebulae}

\author[Jacoby et al.]{
George H. Jacoby,$^{1}$\thanks{E-mail: jacoby@noao.edu}
Todd C. Hillwig,$^{2}$
and David Jones$^{3,4}$
\\
$^{1}$NSF's NOIRLab, 950 N. Cherry Ave., Tucson, AZ 85719, USA\\
$^{2}$Department of Physics and Astronomy, Valparaiso University, Valparaiso, IN 46383, USA\\
$^{3}$Insituto de Astrof\'isica de Canarias, E-38205 La Laguna, Tenerife, Spain\\
$^{4}$Departamento de Astrof\'isica, Universidad de La Laguna, E-38206 La Laguna, Tenerife, Spain
}

\date{Accepted XXX. Received YYY; in original form ZZZ}

\pubyear{2020}

\begin{document}
\label{firstpage}
\pagerange{\pageref{firstpage}--\pageref{lastpage}}
\maketitle

\begin{abstract}
{Eight} planetary nebulae have been identified as `born-again', a class of object typified by knotty secondary ejecta having low masses ($\sim$$10^{-4}$ M$_{\odot}$) with nearly no hydrogen. Abell~30, the archetype of the class, also belongs to a small subset of planetary nebulae that exhibit extreme abundance discrepancy factors (where Abell~30 is the most extreme), a phenomenon strongly linked to binary star interactions. We report the presence of light curve brightness variations having a period of 1.060 days that are highly suggestive of a binary central star in Abell~30. If confirmed, this detection supports the proposed link between binary central stars and extreme abundance discrepancies.
\end{abstract}




\section{Introduction} \label{sec:intro}

\subsection{History of Abell~30 as a born-again PN}
Abell~30 (A30) was first identified on the Palomar Observatory Sky Survey plates by \citet{abell66}. During a narrow-band imaging survey of faint Abell planetary nebulae (PNe), \citet{jacoby79} noted unusual central structures in both A30 and Abell~78 (A78) that are visible in \oiii\  but not in H$\alpha$. \citet{hazard80} also identified central knots in A30 and described additional anomalies that tend to be common across the born-again class.

A30 has since been the focus of numerous studies to assess its chemical composition, photoionization mechanisms, and origin \citep{jacoby83}. { To explain A30 and A78,} \citet{iben83} first proposed a theoretical basis for these born-again central stars in which the star undergoes a late helium flash, often referred to as a very late thermal pulse (VLTP), to explain the hydrogen-deficient ejecta that signal a born-again PN. { \citet{harrington96} proposed an alternative concept requiring binary star interactions.}

{ The following PNe are usually included in the born-again group: A30 \citep{iben83}}, { Abell~58 (A58) \citep{seitter87}}, { A78 \citep{iben83}}, GJJC-1 { (also known as IRAS~18333-2357)} \citep{gillett89}, Sakurai's Object \citep{asplund97}, WR-72 \citep{gvaramadze20}, IRAS~15154-5258 \citep{zijlstra02}, and { HuBi~1 \citep{guerrero18}}. 

\subsection{Connection to nova-like events}
While the VLTP interpretation for born-again PNe is attractive, several objects exhibit strong violations of the chemical composition predictions {\citep{lau11}}. {Specifically, the high Ne/O ($>$0.5, by number) and low C/O ($\ll$1.0) are far more consistent with an ONeMg nova}. For further discussions on the nova hypothesis, see \citet{lau11} for A58, \citet{wesson03} for A30, \citet{jacoby17} for GJJC-1, and \citet{hinkle08} who notes the common properties across objects in this class, including Sakarai and A78, suggesting they have similar origins.

Of special interest here, \citet{lau11} summarize alternatives to the now conventional born-again scenario, but all pathways require binary central stars that have not previously been identified.

\subsection{ADFs and the connection to binarity}

{For over seventy years, it has been known that empirically derived abundances of photoionised nebulae vary significantly depending on whether they are derived using recombination lines or collisionally-excited lines \citep{wyse42}.  The ratio of abundances of the same species derived from recombination lines to those from collisionally-excited lines is known as the abundance discrepancy factor (ADF), and has a median value of 1.9 in H~\textsc{ii} regions and 2.5 in PNe \citep{wesson18}}.  

{Several PNe display ADFs far in excess of this median value \citep[with the knots of A30 having the largest ADF known at almost 1000;][]{wesson03}, a significant number of which have been found to host close-binary central stars \citep{corradi15,jones16}.  Indeed,}
\citet{wesson18} find that the most extreme abundance discrepancies (cases where the O$^{2+}$ ADF $\gtrsim$10) are associated with the shortest period post-common-envelope central stars (P$_\mathrm{orb}<$1.15~d).  

Of the top ten largest ADF PNe known\footnote{See the literature compilation maintained by Roger Wesson at \url{https://www.nebulousresearch.org/adfs/}.}, seven are hosts to post-common-envelope binary central stars, {one (NGC~1501) hosts an H-deficient [WCE] central star, a spectral type that is strongly linked to the born-again scenario \citep{werner99, demarco08}}, and the final two (occupying the number one and three spots, respectively) are A30 and A58. 

\subsection{Evidence for binarity among born-again PNe}
Detecting companions to PN central stars is observationally challenging \citep{jones17,boffin19}. It is especially difficult with ground-based facilities due to complications with weather, scheduling, cadence, and diurnal and annual cycles. The Kepler and K2 missions avoid these issues and are able to identify signatures of binarity, even for systems having very small amplitudes (e.g., $<0.5$ per cent) arising from non-optimal inclination angles, small companions, or distant companions \citep{demarco15}. Even with K2, though, the problem is compounded for PNe where the nebula may introduce a complex and bright `sky' background.

Although the Kepler/K2 missions did not specifically target the born-again group, A30 was observed in K2 Campaign 16. Both, its variability period and amplitude present obstacles for detection from the ground. 

We are not aware of any definitive detections of binarity among the other members of the born-again group. \citet{demarco04}, however, did find the central star of A78 to be a radial velocity variable, with 11 observations showing shifts of up to 23~km~s$^{-1}$ and an rms scatter of 5.1~km~s$^{-1}$. The scatter is roughly twice the average precision of the individual measurements. Unfortunately, with so few observations, it was not possible to rule out wind variability as the origin of the radial velocity variations.

\section{Observations}

\subsection{K2}

\begin{figure}
\centering
\includegraphics[width=0.473\textwidth,trim= 0 13 0 13, clip]{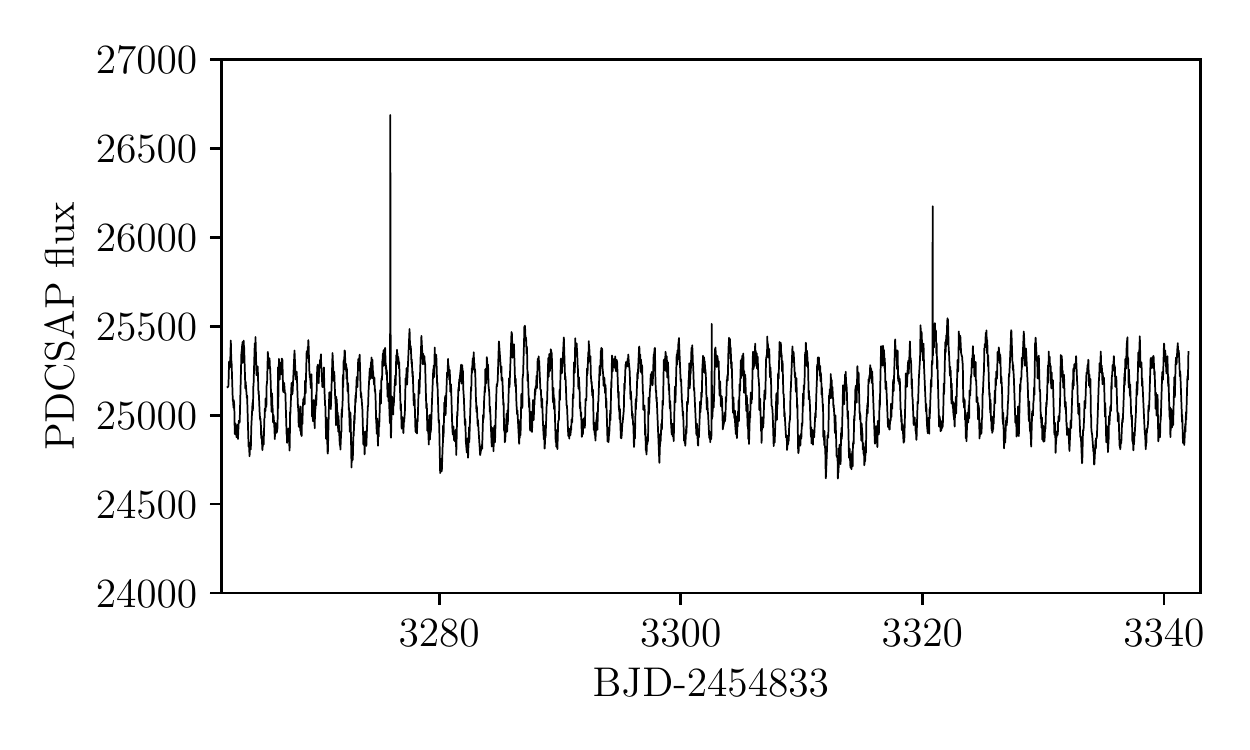}
\caption{The K2 light curve as extracted from the Mikulski Archive for Space Telescopes (MAST). The $\sim$1 day periodicity is evident in the light curve even though the amplitude is only $1.7$ per cent. The curve exhibits several small amplitude residual variations {and two spikes that are data artifacts}, even after the pipeline processing, as is typical for many K2 targets.}
\label{fig:k2_lc}
\end{figure}

A30 was observed for 80 days in the K2 Campaign 16 (2017 December 07 to 2018 February 25) as part of GO programs 16071 (Boyd) and 16901 (Lewis). The pipelined, detrended, PDCSAP light curve is shown in Figure~\ref{fig:k2_lc}. A rapid periodic variability is evident. We also see some residual broad low-amplitude bumps in the curve that survived the detrending processing.

\begin{figure}
\centering
\includegraphics[width=0.473\textwidth]{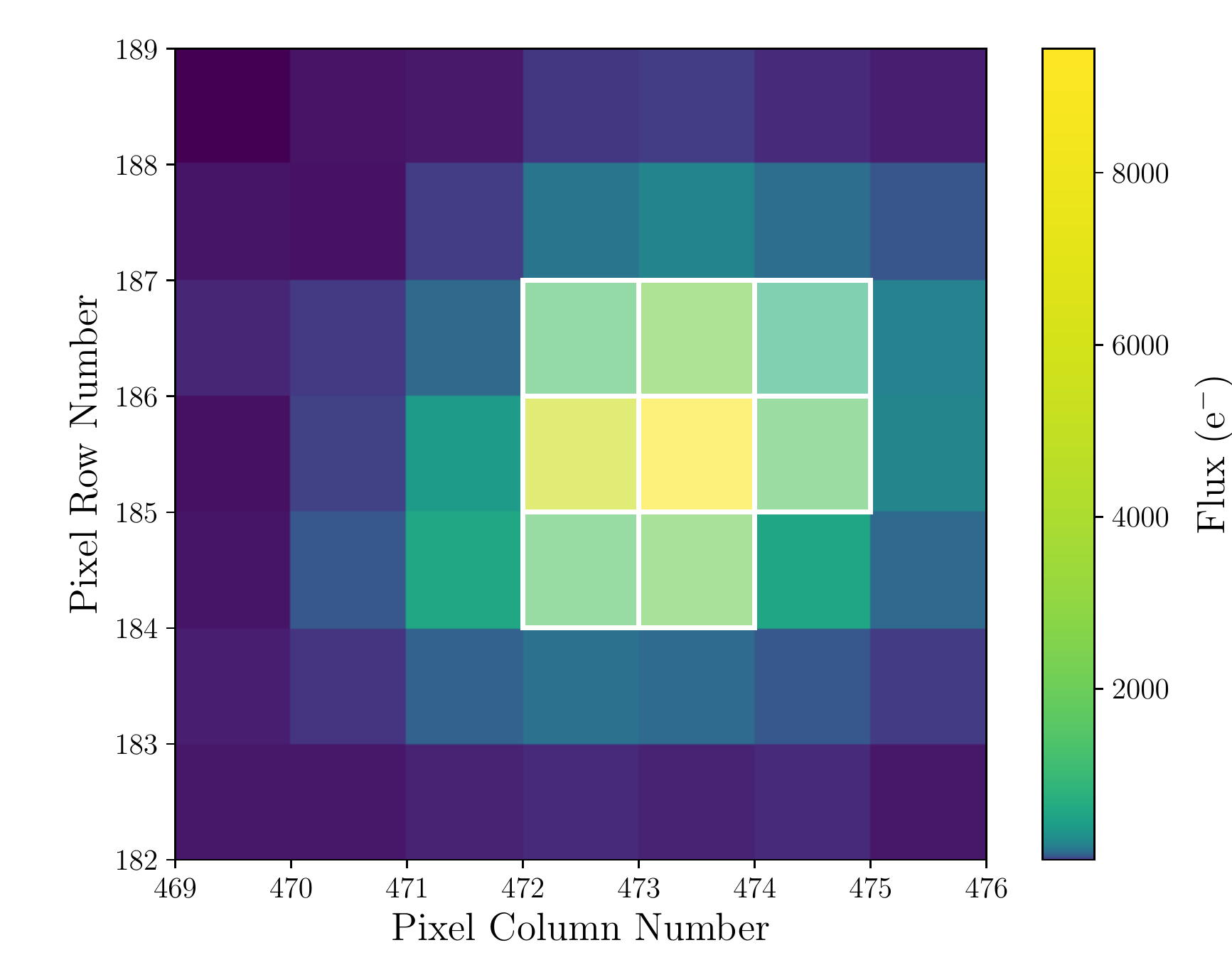}
\caption{The average K2 target pixel file of the central star of A30 with the pixels included in the photometry aperture outlined in white \citep[plot created with an adapted version of \textsc{tpfplotter};][]{aller20}.}
\label{fig:k2_image}
\end{figure}

Figure~\ref{fig:k2_image} shows the K2 extraction region for the central star of A30. Pixels outlined in white represent the mask for the PDCSAP aperture photometry. There is a nearby star \citep{ciardullo99} to the southeast that is 5.2 arcsec away from the central star. Because Kepler/K2 pixels are 3.98 arcsec on a side, the neighbouring star is only 1 pixel away along the diagonal to the lower left.

\begin{figure}
\centering
\includegraphics[width=0.473\textwidth,trim= 0 13 0 13, clip]{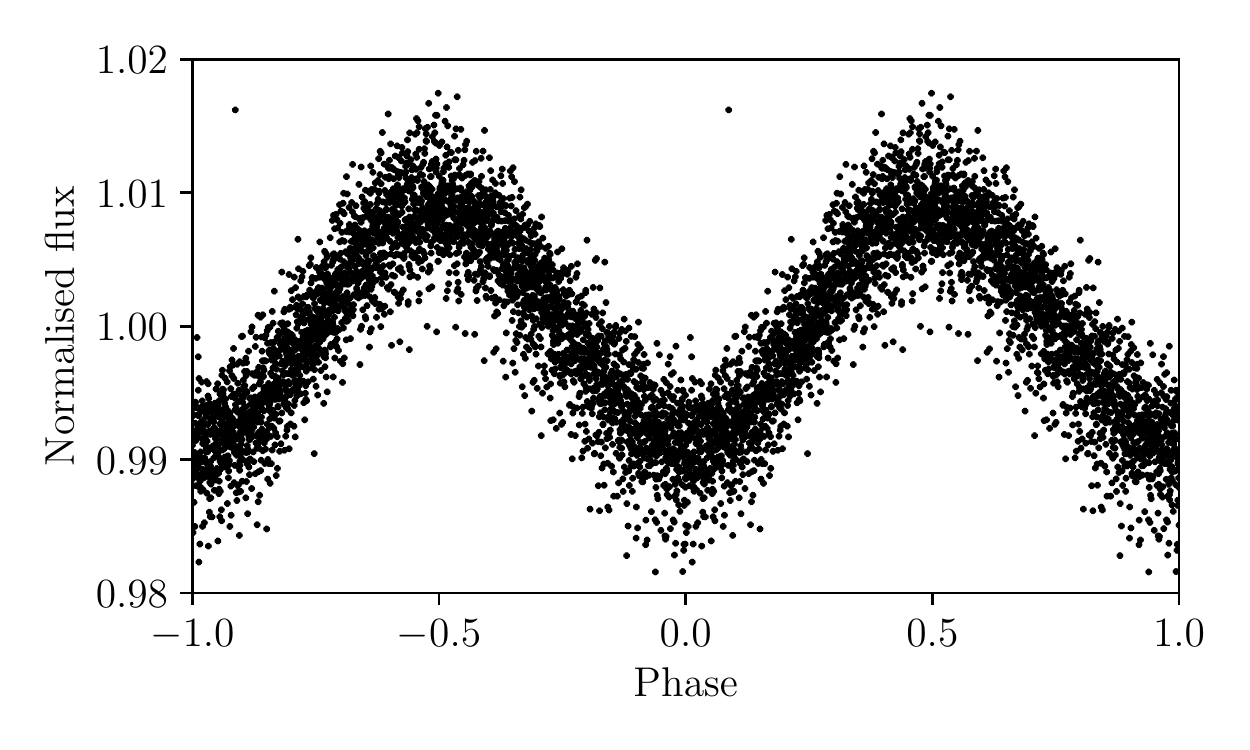}
\caption{The K2 light curve for A30 phased to a period of 1.060 days and normalized to an average of 1.00. Data points have not been either smoothed or averaged.}
\label{fig:k2_phased}
\end{figure}

Using the Lomb-Scargle method \citep[see][]{vanderplas18}, a periodogram of the K2 light curve reveals a strong periodic signal at $1.060\pm0.003$ days, with a peak-to peak-amplitude of $\sim$1.7 per cent. Figure~\ref{fig:k2_phased} shows the light curve phased on the ephemeris of
\begin{equation}
\mathrm{BJD}_\mathrm{min}=245\,8096.25(2) +  1.060(3) E
\end{equation}
for the barycentric Julian date of the photometric minimum ($\mathrm{BJD}_\mathrm{min}$).

\subsection{Other photometric data}
\subsubsection{PanSTARRS}

Because the K2 measurement aperture includes light from the nearby southeast star, it likely contributes to the photometry of A30 in a small way. Although it is 1.7 mag ($\sim$5X) fainter than the A30 central star, it is possible that the neighbour star is the true variable. If so, it would have an intrinsic amplitude 5X larger, or $\sim$9 per cent. 

Fortunately, there are PanSTARRS data for this field with roughly 10 samples over 4.7 years. Based on relative photometry of the neighbouring star, we find that any variations are no larger than $\sim$2 per cent, assuming that we were not unlucky with the PanSTARRS sampling of the phase.

In principle, we can also perform a check on the central star using the same PanSTARRS data. In fact, the sample size of 10 epochs, combined with the small amplitude of 1.7 per cent, does not allow us to conclude that the central star is, or is not, varying. The PanSTARRS photometric quality is quite good, with errors $\sim0.007$ mag, but that is still not sufficiently discriminating when searching for a peak-to-peak amplitude only two times larger with sparse sampling. The problem is exacerbated by variations in seeing across the data set that make accurate and uniform photometry of the central stars of relatively bright and irregular nebulae, like A30, rather challenging \citep{miszalski11,jones15}.

\subsubsection{ATLAS}
We also reviewed data from the Asteroid Terrestrial-impact Last Alert System (ATLAS) sky survey telescope \citep{tonry18}. ATLAS observes the sky every few nights in two filters: cyan ({420-650 nm}) and orange ({560-820 nm}). The cyan data  include 164 samples over 551 days and the orange data  include 153 samples over 543 days. We analysed the two data sets individually, and in combination after applying a color offset.

No periodic variations were detected, likely because the amplitude is quite small and comparable to the ATLAS error bars. Despite the fairly large number of observations, the density of observations was too sparse to detect a nearly 1-day period.

\subsubsection{IAC80}

A30 was observed using the CAMELOT instrument mounted on the IAC80 telescope on the nights of 2005 February 27, and March 17, 26, and  27. On each night, 30--40 contiguous exposures of 120~s were taken using an SDSS-g band filter.  While the observations are insufficient to demonstrate and/or constrain any periodic variations, differential photometry of the central star indicates variability at the 0.03 mag level which phases consistently with the period determined from the K2 data.

Just as for the PanSTARRS data, the IAC80 data allow a further check on the possible variability of the nearby star.  Unfortunately, the relatively large seeing variations (from 1 arcsec to almost 3 arcsec) across the data set limit their photometric value.  However, careful aperture photometry indicates that the nearby star shows little evidence for the $\sim$0.1 mag variability required to explain the observed K2 light curve.

\subsection{Central star spectroscopy}

\subsubsection{VLT-UVES}
\label{sec:uves}

The central star of A30 was observed twice using the Ultraviolet and Visual \'Echelle Spectrograph \citep[UVES;][]{dekker00} mounted on the 8.2-m Very Large Telescope (VLT) of the European Southern Observatory (ESO) as part of the ESO supernovae type Ia progenitor survey \citep[SPY;][]{napiwotzki19}. Observations were acquired on the nights of 2003 January 23 and February 18 using a 2.1 arcsec wide slit and the \#1 dichroic. The red arm of the spectrograph was centred at 564 nm (employing a SHP700 order-blocking filter) and the blue arm was centred at 390 nm (employing an HER5 order-blocking filter) yielding a resolution of approximately 20\,000 in both arms. The pipeline-reduced spectra were downloaded from the ESO archive.

The spectra were analysed by \citet{werner04} who included the central star of A30 in a study of PG 1159 stars. They identified a strong \ion{O}{VI} emission doublet having broad wings ($\sim$1900 km-s$^{-1}$), as well as numerous absorption lines due to \ion{Ne}{VII}.  Also visible in the spectra are strong interstellar Ca H \& K absorption lines.  We used the Ca H \& K lines to define a zero point in radial velocity for the two spectra and analysed the relative radial velocities of the two existing UVES spectra using the \ion{Ne}{VII} lines. We found a relative radial velocity between the two spectra of $-8.1\pm4.6$ km s$^{-1}$, which does not demonstrate clear radial velocity variability.  The period uncertainty and difference in epochs of the UVES and K2 data do not allow us to determine at what point in the photometric variability the spectra were taken. However, using our period of 1.060 days, we find the two spectra were obtained with a phase difference of 0.46. If the system is a binary then either the spectra were taken very near phases 0.0 and 0.5, where the radial velocities would be nearly the same, or the binary inclination is very low.


\subsubsection{WHT-ISIS}

The central star of A30 was observed using the blue arm of the Intermediate-dispersion Spectrograph and Imaging System (ISIS) mounted on the 4.2~m William Herschel Telescope (WHT) on the nights of 2016 January 15 and 16, and February 16. A 0.7 arcsec wide longslit was used along with the R600B grating to provide a resolution of $\sim$1.5~\AA{} over the range of roughly 3600--5100~\AA{}.  All exposures were of 1800~s duration and aligned along the axis of the jet complexes (PA=40$^\circ$ for the January observations and PA=12$^\circ$ for the February observations).

The spacing of the observations, combined with the uncertainty on the photometric ephemeris, means that it is not possible to reliably determine the relative orbital phases of the spectra.  However, based on spectra taken on the same night, we did not find evidence for radial velocity shifts greater than $\sim$30~km~s$^{-1}$ over the course of roughly 0.1 in phase.

\section{Discussion}

It is clear that the K2 observations identified a very significant 1.060 d periodicity in the light curve of the central star of A30.  The consistent amplitude and periodicity essentially rule out wind variability. However, such low-amplitude sinusoidal variability could be a signpost of several other physical processes.

\subsection{Pulsations}

The central star of A30 has been identified as a  [WC]-PG~1159 star, in transition between the [WC] and PG~1159 classes \citep[e.g.;][]{demarco08b}.  PG~1159 stars frequently exhibit low-amplitude \citep[2--20 per cent;][]{grauer87,obrien00}, multi-periodic, photometric variability due to non-radial g-mode pulsations \citep{althaus10}.  Their observed periods lie roughly in the range 300--5\,000~s, and there is theoretical support for pulsation periods up to 10\,000~s \citep{quiron07}.  Thus, the 1.060 day photometric period of A30 is roughly an order of magnitude too long to be attributed to typical PG~1159 pulsations.

\subsection{Star spots}

Precision photometry has begun to reveal that a significant number of hot stars exhibit magnetic spots on their surfaces that result in rotational variability \citep{brinkworth05,reding18,balona19}.  If a spot is responsible for the observed variability of the central star of A30, then the spot must cover a significant fraction of the stellar surface; otherwise the brightness would not be changing continuously over the entire period, as seen by the smooth sinusoidal morphology of the light curve \citep{kilic15,momany20}.

Similarly, the spot would need to be be rather long-lived (both in size and location on the stellar surface), with the K2 variability being constant in amplitude, morphology, and phasing for the full 80~day coverage of the observations.  A recent study of extreme horizontal branch {(EHB)} stars in globular clusters has revealed that the spots on these stars satisfy both these requirements, and are likely maintained by low-level, dynamo-generated magnetic fields \citep{momany20}.  { However, while EHB stars likely have similar variability periods and stellar radii, the temperatures (20--30 kK) and masses $\sim0.48$ M$_\odot$ \citep{momany20} are unlike the hotter and typically more massive central stars of PNe.} It is thus unclear whether similar processes could occur in much hotter stars, like the central star of A30, but we cannot reject the star spot hypothesis.

\subsection{Ellipsoidal modulations}

Tidal distortions in close binaries can lead to photometric variability as the projected surface area of one (or both) components changes as a function of orbital phase \citep[e.g.;][]{hillwig10,santander-garcia15}.  In this case, the orbital period would be double the observed photometric period (with two minima per orbit).  However, this hypothesis can be safely ruled out in A30. \citet{werner04} found a radius of $\sim$$0.2$ R$_\odot$ for the central star. But for an orbital period of 2.12~d, the stellar radius needs to be $>$1~R$_\odot$ for the central star to come close enough to filling its Roche lobe that the distortion produces the observed photometric variability.

\subsection{Doppler beaming}

Relativistic effects, as a result of binary orbital motion, can lead to low-level sinusoidal variations in brightness known as Doppler beaming \citep{shporer10}.  Indeed, one close binary central star was shown to exhibit such variability in its Kepler light curve \citep{demarco15}.  For A30, we can reject this potential cause of the observed variability due to the large amplitude involved relative to that expected for this process. A 2 per cent photometric variation is roughly an order of magnitude larger than can physically be expected from a reasonable range of parameters for this type of system.

\subsection{Irradiation}

Due to the high temperature ($\sim$110\,000 K, \citet{werner04}) of the central star of A30, it could irradiate a lower temperature companion\footnote{In general, irradiation effects are associated with less-evolved (i.e.\ main sequence) companions.  However, it may be possible to observe similar (low-amplitude) effects in double-degenerate systems with large temperature differences \citep{palma05}.}.  The differing projection of the irradiated, or `day', hemisphere of the companion would lead to the observed periodic brightness variations.  The amplitude of the effect is a complex function of the orbital period and inclination, the stellar radii, and the surface temperatures \citep{demarco08,jones17}.  However, it is clear that for the observed amplitude, the irradiated companion would need to be small (perhaps sub-stellar), and/or the binary orbital inclination would have to be close to the plane of the sky. 

Studies of other post-common-envelope binary central stars of PNe indicate that the binary orbital planes are \emph{always} aligned perpendicular to the symmetry axis of the host nebula \citep{hillwig16}.  For A30, the orientation of the hydrogen-poor material surrounding the central star implies a binary orbital inclination of roughly 45$^\circ$ (assuming a flat disk), suggesting the presence of a very small companion. 

{ Roughly 12--15\% of all PN central stars show photometric variability due to a close binary star, most of those dominated by irradiation effects \citep[e.g.][]{miszalski09}.  Star spots have been suggested as a possible, though unlikely, explanation for photometric variability in only one PN central star, that of Pa 5 \citep{demarco15}. Consequently,} we believe that irradiation of a cooler companion is { the most} likely origin for the observed photometric variability in the A30 system.

\section{Conclusions}

We present strong photometric evidence that the central star of A30 has a binary companion with a period of 1.060~days that is being irradiated by the hot central star. However, we are not able to demonstrate a consistent radial velocity variation with the limited phase coverage available from archival spectroscopy. This leaves open the possibility that the photometric variability may not be binary in origin but rather rotational due to a large spot on the surface.  Further observations, particularly high-resolution, time-resolved spectroscopy capable of detecting low-amplitude radial velocity variations and/or mapping the stellar surface via Doppler tomography, will be critical in confirming (or refuting) the binary nature of A30's central star.

These results may constitute an important step in understanding the born-again phenomenon, particularly considering the growing links between high abundance discrepancy planetary nebulae and their nova-like characteristics with close binary evolution.  We encourage observers to obtain precise radial velocities and photometry for the central stars of other members of the born-again class in order to further constrain the importance of binarity in these objects.

\section*{Acknowledgements}

The authors thank Paulina Sowicka for useful discussions on the pulsational properties of PG~1159 stars. {We also thank the anonymous referee for valuable suggestions to improve the presentation.}

This paper uses data from the Kepler and Hubble Space Telescope missions and obtained from the MAST data archive at the Space Telescope Science Institute (STScI). Funding for the Kepler mission is provided by the NASA Science Mission Directorate. STScI is operated by the Association of Universities for Research in Astronomy, Inc., under NASA contract NAS 5–26555. We used \textsc{lightkurve}, a Python package for Kepler and TESS data analysis \citep{lightkurve}. Based on observations made with ESO Telescopes at the La Silla Paranal Observatory under programme ID 167.D-0407.  Based on observations made with the IAC80 operated by the Instituto de Astrof\'isica de Canarias at Observatorio del Teide, and made with the William Herschel Telescope operated by the Isaac Newton Group at the Observatorio del Roque de los Muchachos of the Instituto de Astrofísica de Canarias.  We used the HASH PN database at hashpn.space \citep{parker16}.  The Pan-STARRS1 (PS) Surveys were made possible through contributions of the University of Hawaii, the PS Project, the Max-Planck (MP) Society and institutes, the MP Institutes for Astronomy, Heidelberg and Garching, these universities: Johns Hopkins, Durham, Edinburgh, Queen's Belfast, the Harvard-Smithsonian CfA, Maryland, and Eotvos Lorand, and the Las Cumbres Observatory Global Telescope Network Incorporated, the National Central University of Taiwan, the STScI, NASA Grant NNX08AR22G, and National Science Foundation Grant AST-1238877. 

DJ acknowledges support from the State Research Agency (AEI) of the Spanish Ministry of Science, Innovation and Universities (MCIU) and the European Regional Development Fund (FEDER) under grant AYA2017-83383-P.

\section*{Data availability}
All key data are available from public archives. All secondary data sets, which are not publicly archived, are available upon request.





\bibliographystyle{mnras}
\bibliography{A30} 

\begin{thebibliography}{}
\makeatletter
\relax
\def\mn@urlcharsother{\let\do\@makeother \do\$\do\&\do\#\do\^\do\_\do\%\do\~}
\def\mn@doi{\begingroup\mn@urlcharsother \@ifnextchar [ {\mn@doi@}
  {\mn@doi@[]}}
\def\mn@doi@[#1]#2{\def\@tempa{#1}\ifx\@tempa\@empty \href
  {http://dx.doi.org/#2} {doi:#2}\else \href {http://dx.doi.org/#2} {#1}\fi
  \endgroup}
\def\mn@eprint#1#2{\mn@eprint@#1:#2::\@nil}
\def\mn@eprint@arXiv#1{\href {http://arxiv.org/abs/#1} {{\tt arXiv:#1}}}
\def\mn@eprint@dblp#1{\href {http://dblp.uni-trier.de/rec/bibtex/#1.xml}
  {dblp:#1}}
\def\mn@eprint@#1:#2:#3:#4\@nil{\def\@tempa {#1}\def\@tempb {#2}\def\@tempc
  {#3}\ifx \@tempc \@empty \let \@tempc \@tempb \let \@tempb \@tempa \fi \ifx
  \@tempb \@empty \def\@tempb {arXiv}\fi \@ifundefined
  {mn@eprint@\@tempb}{\@tempb:\@tempc}{\expandafter \expandafter \csname
  mn@eprint@\@tempb\endcsname \expandafter{\@tempc}}}

\bibitem[\protect\citeauthoryear{{Abell}}{{Abell}}{1966}]{abell66}
{Abell} G.~O.,  1966, \mn@doi [\apj] {10.1086/148602}, \href
  {https://ui.adsabs.harvard.edu/abs/1966ApJ...144..259A} {144, 259}

\bibitem[\protect\citeauthoryear{{Aller}, {Lillo-Box}, {Jones}, {Miranda}  \&
  {Barcel{\'o} Forteza}}{{Aller} et~al.}{2020}]{aller20}
{Aller} A.,  {Lillo-Box} J.,  {Jones} D.,  {Miranda} L.~F.,   {Barcel{\'o}
  Forteza} S.,  2020, \mn@doi [\aap] {10.1051/0004-6361/201937118}, \href
  {https://ui.adsabs.harvard.edu/abs/2020A&A...635A.128A} {635, A128}

\bibitem[\protect\citeauthoryear{{Althaus}, {C{\'o}rsico}, {Isern}  \&
  {Garc{\'\i}a-Berro}}{{Althaus} et~al.}{2010}]{althaus10}
{Althaus} L.~G.,  {C{\'o}rsico} A.~H.,  {Isern} J.,   {Garc{\'\i}a-Berro} E.,
  2010, \mn@doi [\aapr] {10.1007/s00159-010-0033-1}, \href
  {https://ui.adsabs.harvard.edu/abs/2010A&ARv..18..471A} {18, 471}

\bibitem[\protect\citeauthoryear{{Asplund}, {Gustafsson}, {Lambert}  \&
  {Kameswara Rao}}{{Asplund} et~al.}{1997}]{asplund97}
{Asplund} M.,  {Gustafsson} B.,  {Lambert} D.~L.,   {Kameswara Rao} N.,  1997,
  \aap, \href {https://ui.adsabs.harvard.edu/abs/1997A&A...321L..17A} {321,
  L17}

\bibitem[\protect\citeauthoryear{{Balona}}{{Balona}}{2019}]{balona19}
{Balona} L.~A.,  2019, \mn@doi [\mnras] {10.1093/mnras/stz2808}, \href
  {https://ui.adsabs.harvard.edu/abs/2019MNRAS.490.2112B} {490, 2112}

\bibitem[\protect\citeauthoryear{{Boffin} \& {Jones}}{{Boffin} \&
  {Jones}}{2019}]{boffin19}
{Boffin} H. M.~J.,  {Jones} D.,  2019, {The Importance of Binaries in the
  Formation and Evolution of Planetary Nebulae}.
{Springer Nature}, \mn@doi{10.1007/978-3-030-25059-1}

\bibitem[\protect\citeauthoryear{{Brinkworth}, {Marsh}, {Morales-Rueda},
  {Maxted}, {Burleigh}  \& {Good}}{{Brinkworth} et~al.}{2005}]{brinkworth05}
{Brinkworth} C.~S.,  {Marsh} T.~R.,  {Morales-Rueda} L.,  {Maxted} P.~F.~L.,
  {Burleigh} M.~R.,   {Good} S.~A.,  2005, \mn@doi [\mnras]
  {10.1111/j.1365-2966.2005.08649.x}, \href
  {https://ui.adsabs.harvard.edu/abs/2005MNRAS.357..333B} {357, 333}

\bibitem[\protect\citeauthoryear{{Ciardullo}, {Bond}, {Sipior}, {Fullton},
  {Zhang}  \& {Schaefer}}{{Ciardullo} et~al.}{1999}]{ciardullo99}
{Ciardullo} R.,  {Bond} H.~E.,  {Sipior} M.~S.,  {Fullton} L.~K.,  {Zhang}
  C.~Y.,   {Schaefer} K.~G.,  1999, \mn@doi [\aj] {10.1086/300940}, \href
  {https://ui.adsabs.harvard.edu/abs/1999AJ....118..488C} {118, 488}

\bibitem[\protect\citeauthoryear{{Corradi}, {Garc{\'\i}a-Rojas}, {Jones}  \&
  {Rodr{\'\i}guez-Gil}}{{Corradi} et~al.}{2015}]{corradi15}
{Corradi} R. L.~M.,  {Garc{\'\i}a-Rojas} J.,  {Jones} D.,
  {Rodr{\'\i}guez-Gil} P.,  2015, \mn@doi [\apj] {10.1088/0004-637X/803/2/99},
  \href {https://ui.adsabs.harvard.edu/abs/2015ApJ...803...99C} {803, 99}

\bibitem[\protect\citeauthoryear{{De Marco}}{{De Marco}}{2008}]{demarco08b}
{De Marco} O.,  2008, in {Werner} A.,  {Rauch} T.,  eds,  ASP Conference Series
  Vol. 391, Hydrogen-Deficient Stars. p.~209 (\mn@eprint {arXiv} {0711.2461})

\bibitem[\protect\citeauthoryear{{De Marco}, {Bond}, {Harmer}  \&
  {Fleming}}{{De Marco} et~al.}{2004}]{demarco04}
{De Marco} O.,  {Bond} H.~E.,  {Harmer} D.,   {Fleming} A.~J.,  2004, \mn@doi
  [\apjl] {10.1086/382156}, \href
  {https://ui.adsabs.harvard.edu/abs/2004ApJ...602L..93D} {602, L93}

\bibitem[\protect\citeauthoryear{{De Marco}, {Hillwig}  \& {Smith}}{{De Marco}
  et~al.}{2008}]{demarco08}
{De Marco} O.,  {Hillwig} T.~C.,   {Smith} A.~J.,  2008, \mn@doi [\aj]
  {10.1088/0004-6256/136/1/323}, \href
  {https://ui.adsabs.harvard.edu/abs/2008AJ....136..323D} {136, 323}

\bibitem[\protect\citeauthoryear{{De Marco}, {Long}, {Jacoby}, {Hillwig},
  {Kronberger}, {Howell}, {Reindl}  \& {Margheim}}{{De Marco}
  et~al.}{2015}]{demarco15}
{De Marco} O.,  {Long} J.,  {Jacoby} G.~H.,  {Hillwig} T.,  {Kronberger} M.,
  {Howell} S.~B.,  {Reindl} N.,   {Margheim} S.,  2015, \mn@doi [\mnras]
  {10.1093/mnras/stv249}, \href
  {https://ui.adsabs.harvard.edu/abs/2015MNRAS.448.3587D} {448, 3587}

\bibitem[\protect\citeauthoryear{{Dekker}, {D'Odorico}, {Kaufer}, {Delabre}  \&
  {Kotzlowski}}{{Dekker} et~al.}{2000}]{dekker00}
{Dekker} H.,  {D'Odorico} S.,  {Kaufer} A.,  {Delabre} B.,   {Kotzlowski} H.,
  2000, in {Iye} M.,  {Moorwood} A.~F.,  eds,  Society of Photo-Optical
  Instrumentation Engineers (SPIE) Conference Series Vol. 4008, \procspie. pp
  534--545, \mn@doi{10.1117/12.395512}

\bibitem[\protect\citeauthoryear{{Gillett}, {Jacoby}, {Joyce}, {Cohen},
  {Neugebauer}, {Soifer}, {Nakajima}  \& {Matthews}}{{Gillett}
  et~al.}{1989}]{gillett89}
{Gillett} F.~C.,  {Jacoby} G.~H.,  {Joyce} R.~R.,  {Cohen} J.~G.,  {Neugebauer}
  G.,  {Soifer} B.~T.,  {Nakajima} T.,   {Matthews} K.,  1989, \mn@doi [\apj]
  {10.1086/167241}, \href
  {https://ui.adsabs.harvard.edu/abs/1989ApJ...338..862G} {338, 862}

\bibitem[\protect\citeauthoryear{{Grauer}, {Bond}, {Liebert}, {Fleming}  \&
  {Green}}{{Grauer} et~al.}{1987}]{grauer87}
{Grauer} A.~D.,  {Bond} H.~E.,  {Liebert} J.,  {Fleming} T.~A.,   {Green}
  R.~F.,  1987, \mn@doi [\apj] {10.1086/165824}, \href
  {https://ui.adsabs.harvard.edu/abs/1987ApJ...323..271G} {323, 271}

\bibitem[\protect\citeauthoryear{{Guerrero} et~al.,}{{Guerrero}
  et~al.}{2018}]{guerrero18}
{Guerrero} M.~A.,  et~al., 2018, \mn@doi [Nature Astronomy]
  {10.1038/s41550-018-0551-8}, \href
  {https://ui.adsabs.harvard.edu/abs/2018NatAs...2..784G} {2, 784}

\bibitem[\protect\citeauthoryear{{Gvaramadze}, {Kniazev}, {Gr{\"a}fener}  \&
  {Langer}}{{Gvaramadze} et~al.}{2020}]{gvaramadze20}
{Gvaramadze} V.~V.,  {Kniazev} A.~Y.,  {Gr{\"a}fener} G.,   {Langer} N.,  2020,
  \mn@doi [\mnras] {10.1093/mnras/stz3639}, \href
  {https://ui.adsabs.harvard.edu/abs/2020MNRAS.492.3316G} {492, 3316}

\bibitem[\protect\citeauthoryear{{Harrington}}{{Harrington}}{1996}]{harrington96}
{Harrington} J.~P.,  1996, in {Jeffery} C.~S.,  {Heber} U.,  eds,  ASP
  Conference Series Vol. 96, Hydrogen Deficient Stars. p.~193

\bibitem[\protect\citeauthoryear{{Hazard}, {Terlevich}, {Morton}, {Sargent}  \&
  {Ferland}}{{Hazard} et~al.}{1980}]{hazard80}
{Hazard} C.,  {Terlevich} R.,  {Morton} D.~C.,  {Sargent} W.~L.~W.,   {Ferland}
  G.,  1980, \mn@doi [\nat] {10.1038/285463a0}, \href
  {https://ui.adsabs.harvard.edu/abs/1980Natur.285..463H} {285, 463}

\bibitem[\protect\citeauthoryear{{Hillwig}, {Bond}, {Af{\textcommabelow s}ar}
  \& {De Marco}}{{Hillwig} et~al.}{2010}]{hillwig10}
{Hillwig} T.~C.,  {Bond} H.~E.,  {Af{\textcommabelow s}ar} M.,   {De Marco} O.,
   2010, \mn@doi [\aj] {10.1088/0004-6256/140/2/319}, \href
  {https://ui.adsabs.harvard.edu/abs/2010AJ....140..319H} {140, 319}

\bibitem[\protect\citeauthoryear{{Hillwig}, {Jones}, {De Marco}, {Bond},
  {Margheim}  \& {Frew}}{{Hillwig} et~al.}{2016}]{hillwig16}
{Hillwig} T.~C.,  {Jones} D.,  {De Marco} O.,  {Bond} H.~E.,  {Margheim} S.,
  {Frew} D.,  2016, \mn@doi [\apj] {10.3847/0004-637X/832/2/125}, \href
  {https://ui.adsabs.harvard.edu/abs/2016ApJ...832..125H} {832, 125}

\bibitem[\protect\citeauthoryear{{Hinkle}, {Lebzelter}, {Joyce}, {Ridgway},
  {Close}, {Hron}  \& {Andre}}{{Hinkle} et~al.}{2008}]{hinkle08}
{Hinkle} K.~H.,  {Lebzelter} T.,  {Joyce} R.~R.,  {Ridgway} S.,  {Close} L.,
  {Hron} J.,   {Andre} K.,  2008, \mn@doi [\aap] {10.1051/0004-6361:20077738},
  \href {https://ui.adsabs.harvard.edu/abs/2008A&A...479..817H} {479, 817}

\bibitem[\protect\citeauthoryear{{Iben}, {Kaler}, {Truran}  \&
  {Renzini}}{{Iben} et~al.}{1983}]{iben83}
{Iben} I. J.,  {Kaler} J.~B.,  {Truran} J.~W.,   {Renzini} A.,  1983, \mn@doi
  [\apj] {10.1086/160631}, \href
  {https://ui.adsabs.harvard.edu/abs/1983ApJ...264..605I} {264, 605}

\bibitem[\protect\citeauthoryear{{Jacoby}}{{Jacoby}}{1979}]{jacoby79}
{Jacoby} G.~H.,  1979, \mn@doi [\pasp] {10.1086/130582}, \href
  {https://ui.adsabs.harvard.edu/abs/1979PASP...91..754J} {91, 754}

\bibitem[\protect\citeauthoryear{{Jacoby} \& {Ford}}{{Jacoby} \&
  {Ford}}{1983}]{jacoby83}
{Jacoby} G.~H.,  {Ford} H.~C.,  1983, \mn@doi [\apj] {10.1086/160779}, \href
  {https://ui.adsabs.harvard.edu/abs/1983ApJ...266..298J} {266, 298}

\bibitem[\protect\citeauthoryear{{Jacoby}, {De Marco}, {Davies}, {Lotarevich},
  {Bond}, {Harrington}  \& {Lanz}}{{Jacoby} et~al.}{2017}]{jacoby17}
{Jacoby} G.~H.,  {De Marco} O.,  {Davies} J.,  {Lotarevich} I.,  {Bond} H.~E.,
  {Harrington} J.~P.,   {Lanz} T.,  2017, \mn@doi [\apj]
  {10.3847/1538-4357/836/1/93}, \href
  {https://ui.adsabs.harvard.edu/abs/2017ApJ...836...93J} {836, 93}

\bibitem[\protect\citeauthoryear{{Jones} \& {Boffin}}{{Jones} \&
  {Boffin}}{2017}]{jones17}
{Jones} D.,  {Boffin} H. M.~J.,  2017, \mn@doi [Nature Astronomy]
  {10.1038/s41550-017-0117}, \href
  {https://ui.adsabs.harvard.edu/abs/2017NatAs...1E.117J} {1, 0117}

\bibitem[\protect\citeauthoryear{{Jones}, {Boffin}, {Rodr{\'\i}guez-Gil},
  {Wesson}, {Corradi}, {Miszalski}  \& {Mohamed}}{{Jones}
  et~al.}{2015}]{jones15}
{Jones} D.,  {Boffin} H.~M.~J.,  {Rodr{\'\i}guez-Gil} P.,  {Wesson} R.,
  {Corradi} R.~L.~M.,  {Miszalski} B.,   {Mohamed} S.,  2015, \mn@doi [\aap]
  {10.1051/0004-6361/201425454}, \href
  {https://ui.adsabs.harvard.edu/abs/2015A&A...580A..19J} {580, A19}

\bibitem[\protect\citeauthoryear{{Jones}, {Wesson}, {Garc{\'\i}a-Rojas},
  {Corradi}  \& {Boffin}}{{Jones} et~al.}{2016}]{jones16}
{Jones} D.,  {Wesson} R.,  {Garc{\'\i}a-Rojas} J.,  {Corradi} R.~L.~M.,
  {Boffin} H.~M.~J.,  2016, \mn@doi [\mnras] {10.1093/mnras/stv2519}, \href
  {https://ui.adsabs.harvard.edu/abs/2016MNRAS.455.3263J} {455, 3263}

\bibitem[\protect\citeauthoryear{{Kilic} et~al.,}{{Kilic}
  et~al.}{2015}]{kilic15}
{Kilic} M.,  et~al., 2015, \mn@doi [\apjl] {10.1088/2041-8205/814/2/L31}, \href
  {https://ui.adsabs.harvard.edu/abs/2015ApJ...814L..31K} {814, L31}

\bibitem[\protect\citeauthoryear{{Lau}, {De Marco}  \& {Liu}}{{Lau}
  et~al.}{2011}]{lau11}
{Lau} H. H.~B.,  {De Marco} O.,   {Liu} X.~W.,  2011, \mn@doi [\mnras]
  {10.1111/j.1365-2966.2010.17568.x}, \href
  {https://ui.adsabs.harvard.edu/abs/2011MNRAS.410.1870L} {410, 1870}

\bibitem[\protect\citeauthoryear{{Lightkurve Collaboration}
  et~al.,}{{Lightkurve Collaboration} et~al.}{2018}]{lightkurve}
{Lightkurve Collaboration} et~al., 2018, {Lightkurve: Kepler and TESS time
  series analysis in Python}, Astrophysics Source Code Library (\mn@eprint
  {ascl} {1812.013})

\bibitem[\protect\citeauthoryear{{Miszalski}, {Acker}, {Moffat}, {Parker}  \&
  {Udalski}}{{Miszalski} et~al.}{2009}]{miszalski09}
{Miszalski} B.,  {Acker} A.,  {Moffat} A.~F.~J.,  {Parker} Q.~A.,   {Udalski}
  A.,  2009, \mn@doi [\aap] {10.1051/0004-6361/200811380}, \href
  {https://ui.adsabs.harvard.edu/abs/2009A&A...496..813M} {496, 813}

\bibitem[\protect\citeauthoryear{{Miszalski}, {Jones}, {Rodr{\'\i}guez-Gil},
  {Boffin}, {Corradi}  \& {Santander-Garc{\'\i}a}}{{Miszalski}
  et~al.}{2011}]{miszalski11}
{Miszalski} B.,  {Jones} D.,  {Rodr{\'\i}guez-Gil} P.,  {Boffin} H.~M.~J.,
  {Corradi} R.~L.~M.,   {Santander-Garc{\'\i}a} M.,  2011, \mn@doi [\aap]
  {10.1051/0004-6361/201117084}, \href
  {https://ui.adsabs.harvard.edu/abs/2011A&A...531A.158M} {531, A158}

\bibitem[\protect\citeauthoryear{{Momany} et~al.,}{{Momany}
  et~al.}{2020}]{momany20}
{Momany} Y.,  et~al., 2020, \mn@doi [Nature Astronomy]
  {10.1038/s41550-020-1113-4}, \href
  {https://ui.adsabs.harvard.edu/abs/2020NatAs.tmp..116M} {}

\bibitem[\protect\citeauthoryear{{Napiwotzki} et~al.,}{{Napiwotzki}
  et~al.}{2019}]{napiwotzki19}
{Napiwotzki} R.,  et~al., 2019, arXiv e-prints, \href
  {https://ui.adsabs.harvard.edu/abs/2019arXiv190610977N} {p. arXiv:1906.10977}

\bibitem[\protect\citeauthoryear{{O'Brien}}{{O'Brien}}{2000}]{obrien00}
{O'Brien} M.~S.,  2000, \mn@doi [\apj] {10.1086/308613}, \href
  {https://ui.adsabs.harvard.edu/abs/2000ApJ...532.1078O} {532, 1078}

\bibitem[\protect\citeauthoryear{{Palma} \& {Rohrmann}}{{Palma} \&
  {Rohrmann}}{2005}]{palma05}
{Palma} T.,  {Rohrmann} R.~D.,  2005, in {Koester} D.,  {Moehler} S.,  eds,
  ASP Conference Series Vol. 334, 14th European Workshop on White Dwarfs.
  p.~285

\bibitem[\protect\citeauthoryear{{Parker}, {Boji{\v{c}}i{\'c}}  \&
  {Frew}}{{Parker} et~al.}{2016}]{parker16}
{Parker} Q.~A.,  {Boji{\v{c}}i{\'c}} I.~S.,   {Frew} D.~J.,  2016, in Journal
  of Physics Conference Series. p. 032008 (\mn@eprint {arXiv} {1603.07042}),
  \mn@doi{10.1088/1742-6596/728/3/032008}

\bibitem[\protect\citeauthoryear{{Quirion}, {Fontaine}  \&
  {Brassard}}{{Quirion} et~al.}{2007}]{quiron07}
{Quirion} P.~O.,  {Fontaine} G.,   {Brassard} P.,  2007, \mn@doi [\apjs]
  {10.1086/513870}, \href
  {https://ui.adsabs.harvard.edu/abs/2007ApJS..171..219Q} {171, 219}

\bibitem[\protect\citeauthoryear{{Reding}, {Hermes}  \& {Clemens}}{{Reding}
  et~al.}{2018}]{reding18}
{Reding} J.~S.,  {Hermes} J.~J.,   {Clemens} J.~C.,  2018, in Proceedings of
  the 21st European Workshop on White Dwarfs. p.~1, \mn@doi{10.26153/tsw/967}

\bibitem[\protect\citeauthoryear{{Santander-Garc{\'\i}a}, {Rodr{\'\i}guez-Gil},
  {Corradi}, {Jones}, {Miszalski}, {Boffin}, {Rubio-D{\'\i}ez}  \&
  {Kotze}}{{Santander-Garc{\'\i}a} et~al.}{2015}]{santander-garcia15}
{Santander-Garc{\'\i}a} M.,  {Rodr{\'\i}guez-Gil} P.,  {Corradi} R.~L.~M.,
  {Jones} D.,  {Miszalski} B.,  {Boffin} H.~M.~J.,  {Rubio-D{\'\i}ez} M.~M.,
  {Kotze} M.~M.,  2015, \mn@doi [\nat] {10.1038/nature14124}, \href
  {https://ui.adsabs.harvard.edu/abs/2015Natur.519...63S} {519, 63}

\bibitem[\protect\citeauthoryear{{Seitter}}{{Seitter}}{1987}]{seitter87}
{Seitter} W.~C.,  1987, The Messenger, \href
  {https://ui.adsabs.harvard.edu/abs/1987Msngr..50...14S} {50, 14}

\bibitem[\protect\citeauthoryear{{Shporer}, {Kaplan}, {Steinfadt}, {Bildsten},
  {Howell}  \& {Mazeh}}{{Shporer} et~al.}{2010}]{shporer10}
{Shporer} A.,  {Kaplan} D.~L.,  {Steinfadt} J. D.~R.,  {Bildsten} L.,  {Howell}
  S.~B.,   {Mazeh} T.,  2010, \mn@doi [\apjl] {10.1088/2041-8205/725/2/L200},
  \href {https://ui.adsabs.harvard.edu/abs/2010ApJ...725L.200S} {725, L200}

\bibitem[\protect\citeauthoryear{{Tonry} et~al.,}{{Tonry}
  et~al.}{2018}]{tonry18}
{Tonry} J.~L.,  et~al., 2018, \mn@doi [\apj] {10.3847/1538-4357/aae386}, \href
  {https://ui.adsabs.harvard.edu/abs/2018ApJ...867..105T} {867, 105}

\bibitem[\protect\citeauthoryear{{VanderPlas}}{{VanderPlas}}{2018}]{vanderplas18}
{VanderPlas} J.~T.,  2018, \mn@doi [\apjs] {10.3847/1538-4365/aab766}, \href
  {https://ui.adsabs.harvard.edu/abs/2018ApJS..236...16V} {236, 16}

\bibitem[\protect\citeauthoryear{{Werner}, {Dreizler}, {Rauch}, {Koesterke}  \&
  {Heber}}{{Werner} et~al.}{1999}]{werner99}
{Werner} K.,  {Dreizler} S.,  {Rauch} T.,  {Koesterke} L.,   {Heber} U.,  1999,
  in {Le Bertre} T.,  {Lebre} A.,   {Waelkens} C.,  eds,  IAU Symposium Vol.
  191, Asymptotic Giant Branch Stars. p.~493

\bibitem[\protect\citeauthoryear{{Werner}, {Rauch}, {Reiff}, {Kruk}  \&
  {Napiwotzki}}{{Werner} et~al.}{2004}]{werner04}
{Werner} K.,  {Rauch} T.,  {Reiff} E.,  {Kruk} J.~W.,   {Napiwotzki} R.,  2004,
  \mn@doi [\aap] {10.1051/0004-6361:20041165}, \href
  {https://ui.adsabs.harvard.edu/abs/2004A&A...427..685W} {427, 685}

\bibitem[\protect\citeauthoryear{{Wesson}, {Liu}  \& {Barlow}}{{Wesson}
  et~al.}{2003}]{wesson03}
{Wesson} R.,  {Liu} X.~W.,   {Barlow} M.~J.,  2003, \mn@doi [\mnras]
  {10.1046/j.1365-8711.2003.06289.x}, \href
  {https://ui.adsabs.harvard.edu/abs/2003MNRAS.340..253W} {340, 253}

\bibitem[\protect\citeauthoryear{{Wesson}, {Jones}, {Garc{\'\i}a-Rojas},
  {Boffin}  \& {Corradi}}{{Wesson} et~al.}{2018}]{wesson18}
{Wesson} R.,  {Jones} D.,  {Garc{\'\i}a-Rojas} J.,  {Boffin} H.~M.~J.,
  {Corradi} R.~L.~M.,  2018, \mn@doi [\mnras] {10.1093/mnras/sty1871}, \href
  {https://ui.adsabs.harvard.edu/abs/2018MNRAS.480.4589W} {480, 4589}

\bibitem[\protect\citeauthoryear{{Wyse}}{{Wyse}}{1942}]{wyse42}
{Wyse} A.~B.,  1942, \mn@doi [\apj] {10.1086/144409}, \href
  {https://ui.adsabs.harvard.edu/abs/1942ApJ....95..356W} {95, 356}

\bibitem[\protect\citeauthoryear{{Zijlstra}}{{Zijlstra}}{2002}]{zijlstra02}
{Zijlstra} A.~A.,  2002, \apss, \href
  {https://ui.adsabs.harvard.edu/abs/2002Ap&SS.279..171Z} {279, 171}

\makeatother
\end{thebibliography}

\bsp	
\label{lastpage}
\end{document}